\documentclass[aps,prb,twocolumn,showpacs]{revtex4}
\usepackage{amssymb}
\usepackage{graphicx}
\usepackage{amsmath}
\usepackage{amsfonts}
\begin{document}

\title{Possibility of $S=1$ spin liquids with fermionic spinons on triangular lattices}

\author{Zheng-Xin Liu$^{1}$}
\author{Yi Zhou$^{2}$}
\email{yizhou@zju.edu.cn}
\author{Tai-Kai Ng$^{1}$ }

\affiliation{1, Department of Physics, Hong Kong University of
Science and Technology, Clear Water Bay Road, Kowloon, Hong Kong\\
2, Department of Physics, Zhejiang University, Hangzhou 310027, P.
R. China}

\begin{abstract}

 In this paper we generalize the fermionic representation for $S=1/2$ spins to arbitrary spins. Within a mean field theory we
 obtain several spin liquid states for spin $S=1$ antiferromagnets on triangular lattices, including gapless f-wave spin
 liquid and topologically nontrivial $p_x+ip_y$ spin liquid. After considering different competing orders, we construct a
 phase diagram for the $J_1$-$J_3$-$K$ model. The application to recently discovered material $\mathrm{NiGa_2S_4}$ is discussed.

\end{abstract}

 \maketitle

\section{introduction}

 Spin liquids are novel quantum magnetic states where long ranged magnetic order is absent at zero temperature due to strong
 quantum fluctuations\cite{AndersonSL}. Instead of spin wave excitations in spin ordered systems, spinons are proposed to be
 the elementary spin excitations in spin liquids. It is believed that spin liquid states can be found in spin $S=1/2$
 antiferromagnets(AFMs) on geometric frustrated lattices and several promising candidate materials have been experimentally
 discovered\cite {Leescience}. A natural question is whether spin liquid states with fermionic spinons can also exist in
 $S>1/2$ systems as is proposed for $S=1/2$ systems.

To address this issue, we formulate a fully quantum mechanical
fermionic mean field theory for $S=1$ system. We study the
Heisenberg AFM, and obtain spin-liquid type solutions which have
not been proposed previously. We focus our interest on the
$J_1$-$J_3$-$K$ model, which is proposed to be the microscopic
Hamiltonian for the interesting  material $\mathrm{NiGa_2S_4}$, an
frustrated AFM on triangular lattice. We argue that a gapless spin
liquid state obtained in our mean-field theory is a candidate for
the ground state when compared with experimental results.

{\section{Fermionic representation of spin}}

To begin with, we introduce the fermionic representation for
spins. In the $S=1/2$ case, two species of fermionic spinons
representing up and down spins are introduced to construct the
spin operators. This fermionic representation can be generalized
to arbitrary spin\cite{B}, in the present paper, we only consider
the case $S=1$. We introduce $3$ species of spinon operators $c_1,
c_0, c_{-1}$ satisfying anti-commutation relations
$\{c_m,c^\dagger_n\}=\delta_{mn}$, where $m,n=1,0,-1$. It is easy
to show that spin operators can be expressed in terms of $c_m$ and
$c^\dagger_n$'s, $\hat {\mathbf S}=C^\dagger {\mathbf I}C$, where
$C=(c_1, c_{0},c_{-1})^T$ and $I^\alpha (\alpha=x,y,z)$ is a
$3\times3$ matrix whose matrix elements are given by
$I^\alpha_{mn}= \langle m|S^\alpha|n\rangle$.

In this fermionic spinon representation, a constraint has to be
imposed on the Hilbert space to ensure that there is only one
fermion per site (particle representation, $N_f=1$).
Alternatively, a spin can equally be represented in a Hilbert
space with $2$ fermions per site (hole representation, $N_f=2$).
The two representations are identical for $S=1/2$, reflecting a
particle-hole symmetry of the Hilbert space which is absent for
$S=1$. For $S=1$ the two representations are related by a symmetry
group of the spin operators as we shall explain in the following.

Following Affleck {\it et al.}\cite{AZHA-1/2}, we introduce the
``hole'' operators $\bar C=(c^\dagger_{-1}, -c^\dagger_{0},
c^\dagger_{1})^T$. It is easy to check that $\bar C$ and $C$
behave in the same manner under spin rotation and the spin
operators can also be written in terms of $\bar C$: $ \hat
{\mathbf S}=\bar C^\dagger {\mathbf I}\bar C$. Combining $C$ and
$\bar C$ into a $3\times2$ matrix $\psi=(C,\bar C)$
\cite{AZHA-1/2}, we can reexpress the spin operator as
\begin{eqnarray}\label{spin}
\hat {\mathbf S}=\frac{1}{2}Tr(\psi^\dagger {\mathbf I}\psi),
\end{eqnarray}
and the constraints can be represented as
\begin{eqnarray}\label{contr2}
\label{newcontr1}Tr(\psi\sigma_z\psi^\dagger)=3-2N_f=\pm1,
\end{eqnarray}
where $+$sign for ``particle" and $-$sign for ``hole"
representations, respectively.

The spin operator (\ref{spin}) is invariant under certain
transformation of the spinon operators $\psi\to\psi W$. These
transformations $W$ form a $U(1)\bar\otimes Z_2$ group (we note
that for half-integer spins the symmetry group is
$SU(2)$\cite{B}). For $S=1/2$, the constraint is invariant under
the $SU(2)$ group because of particle-hole symmetry mentioned
above. However, for $S=1$ the particle-hole symmetry is absent and
the two constraints in Eq.(\ref{contr2}) are not invariant under
the symmetry group. In fact, the two constraints can be
transformed from one to the other by the particle-hole
transformation. We shall adopt the ``particle" representation
$N_f=1$ in the following discussion.
\begin{figure}[hpbt]
\includegraphics[width=7.2cm]{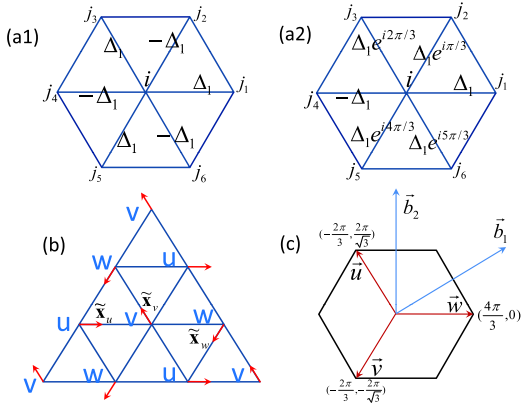}
\caption{(Color online) (a) Two spin liquid ansatzs with different
pairing symmetry. (a1) $f$-wave pairing; (a2) $p_x+ip_y$-wave
pairing. (b) $u$, $v$ and $w$ are the sublattice index. The red
arrows symbol the orientation of the spins and the corresponding
new axes $\tilde x_{u},\tilde x_{v},\tilde x_{w}$. (c) The first
Brillouin zone of triangular lattice.}\label{fig:pairing}
\end{figure}

{\section{The extended Heisenberg model}}

We now apply the fermionic representation to frustrated 2D $S=1$
spin models. We focus on the $J_1$-$J_3$-$K$ model on triangular
lattices,
\begin{eqnarray}\label{Hamiltonian}
H=\sum_{\langle i,j\rangle} \left[ J_1\mathbf  S_i\cdot\mathbf S_j
+K(\mathbf  S_i\cdot\mathbf  S_j)^2\right]+ J_3\sum_{[i,j]}\mathbf
S_i\cdot \mathbf S_{j}.
\end{eqnarray}
where $\langle i,j\rangle$ denotes nearest neighbor (NN) and
$[i,j]$ the third nearest neighbors (NNNN). Several semi-classical
mean field studies of this Hamiltonian have appeared in
literature\cite{AFQ,FQ,Qphase,J1J3K,SU(3)} where most of the trial
ground states are unentangled states (or direct product of local
states). Here we consider a fully quantum mechanical mean field
theory based on the fermion representation which admits resonant
valence bond (RVB) type spin liquid ground states. We first
consider the case $K=0$, $J_1, J_3>0$.

\subsection{Spin liquid solutions at $K=0$}

Similar to the spin-1/2 systems, the following expression also
holds for $S=1$,
\begin{eqnarray}\label{SS}
\mathbf S_i\cdot\mathbf
S_j&=&-{1\over2}Tr:(\psi_j^\dagger\psi_i\psi_i^\dagger\psi_j):
\nonumber\\&=&-:(\chi_{ij}^\dagger\chi_{ij}+\Delta_{ij}^\dagger\Delta_{ij}):
\end{eqnarray}
where $: :$ denotes normal ordering, $\chi_{ij}=C_i^\dagger
C_j=c_{1i}^\dagger c_{1j}+c_{0i}^\dagger c_{0j}+ c_{-1i}^\dagger
c_{-1j}$ is an effective (spin singlet) hopping and $
\Delta_{ij}=\bar C_i^\dagger C_j=c_{-1i}c_{1j}
-c_{0i}c_{0j}+c_{1i}c_{-1j}$ represents ($S=1$) spin-singlet
pairing. A mean field theory can be formulated by replacing one of
the operators by its expectation value, $\mathbf S_i\cdot\mathbf
S_j\sim-(\chi_{ij}^\dagger\langle\chi_{ij}\rangle
+\Delta_{ij}^\dagger\langle\Delta_{ij}\rangle+h.c.)+\langle\chi_{ij}
\rangle^2+\langle\Delta_{ij}\rangle^2$. Notice that
$\langle\Delta_{ji}\rangle=-\langle\Delta_{ij}\rangle$ and the
pairing has odd parity which is different from the corresponding
$S=1/2$ RVB states. We shall first consider solutions which
respect both translational and rotational symmetries. Two such
solutions with $f$-wave and $p_x+ip_y$-wave symmetries
respectively, are obtained\cite{footnote}. The mean field
Hamiltonian of the two states has the form
\begin{figure}[hbpt]
\includegraphics[width=5.5cm]{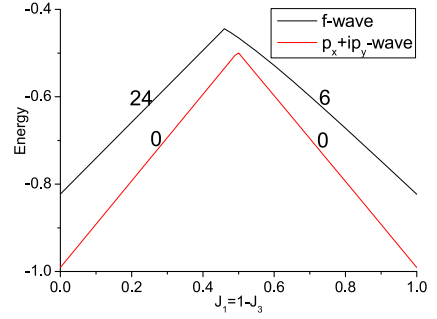}
\caption{ (Color online) A first order phase transition between
two spin liquids occurs at $J_1\sim0.5$. We have set $J_1+J_3=1$.
The number on the each line indicates the number of Dirac cones of
the corresponding state in the first Brillouin zone.
}\label{fig:HSB}
\end{figure}
\begin{eqnarray*}
H_{MF}&=&\sum_{k}\chi_{k}(c_{1k}^\dagger c_{1k}+c_{0k}^\dagger
c_{0k}+c_{-1k}^\dagger c_{-1k})\nonumber\\&&
-\sum_k[\Delta_k(c_{-1-k}c_{1k}-{1\over2}c_{0-k}c_{0k})+h.c.],
\end{eqnarray*}
with $\chi_k=\lambda-Z(J_1\chi_1\gamma_k+J_3\chi_3\gamma_{2k})$,
$\Delta_k=iZ(J_1\Delta_1\psi_k+J_3\Delta_3\psi_{2k})$. Here $Z=6$
is the coordination number, $\lambda$ is the lagrangian multiplier
determined by 
$\langle C_i^\dagger C_i \rangle=1$, and $\gamma_k={1\over3}[\cos
k_x+\cos(-{k_x\over2}+{\sqrt3
k_y\over2})+\cos(-{k_x\over2}-{\sqrt3 k_y\over2})]$.
$\psi_k^{f}={1\over3}[\sin k_x+\sin(-{k_x\over2}+{\sqrt3
k_y\over2})+\sin(-{k_x\over2}-{\sqrt3 k_y\over2})]$ and $\psi_k
^{p_x+ip_y}={1\over3}[\sin k_x+e^{i{\pi\over3}}\sin({k_x\over2}+
{\sqrt3 k_y\over2})+e^{i{2\pi\over3}} \sin(-{k_x\over2}+{\sqrt3
k_y\over2})]$.  The pairing symmetries for these states on the
lattice are illustrated in Fig. \ref{fig:pairing}(a).

The mean field Hamiltonian can be diagonalized with appropriate
Bogoliubov transformations, and the parameters $\chi$, $\Delta$
and $\lambda$ are determined by the self-consistent equations,
\begin{eqnarray}\label{MFeqs}
 \chi_1&=&\langle C_{i}^\dagger C_{i+\mathbf x} \rangle,\ \
 \chi_3=\langle C_{i}^\dagger C_{i+2\mathbf x} \rangle, \nonumber\\
 \Delta_{1}&=&\langle \bar C_i^\dagger C_{i+\mathbf x} \rangle, \ \
 \Delta_{3}=\langle \bar C_i^\dagger C_{i+2\mathbf x}\rangle,\nonumber\\
 1&=&\langle C_i^\dagger C_i\rangle,
\end{eqnarray}
where $i+\mathbf x$ denotes a NN site of $i$ and $i+2\mathbf x$ a
NNNN site along the $x$ direction, $\chi_{1(3)}$ and
$\Delta_{1(3)}$ are parameters on NN(NNNN) bonds. Similar to the
spin-1/2 mean field theory, a physical spin liquid state can be
formed by Gutzwiller projection of the mean field ground state to
the state with single occupancy.

The mean field Hamiltonian describes three branches of fermionic
spinon excitations with $S_z=0,\pm1$ and identical dispersion
$E_k=\sqrt{\chi_k^2+|\Delta_k|^2}$. For the $f$-wave pairing, the
excitation is gapless with several Dirac cones in the Brillouin
zone (the number of cones is given in Fig. \ref{fig:HSB}). For the
$p_x+ip_y$-wave pairing, the bulk excitation is fully gapped.
Since $\chi_k<0$ at the $\Gamma$ point, the $p_x+ip_y$ ansatz
belongs to the weak pairing region\cite{Read}, and there should
exist gapless (chiral) Majorana edge modes on the open boundaries.
Thus the state describes a time-reversal symmetry breaking
topological spin liquid. The $p_x+ip_y$ state has slightly lower
energy in mean-field level.

Our mean field theory predicts two different spin liquid states
(for any fixed pairing symmetry) as a function of $J_1/J_3$. The
$p_x+ip_y$ state remains lower in energy in both cases. A first
order phase transition occurs at $J_1/J_3\sim1$. When $J_1$
dominates, the spin liquid state is characterized by
$\chi_{1,3}\neq0$ and $\Delta_{1,3}\neq0$ (consequently
$\langle\mathbf S_i\cdot\mathbf S_{i+1}\rangle<0$ and
$\langle\mathbf S_i\cdot\mathbf S_{i+2}\rangle<0$); while when
$J_3$ dominates, $\chi_1=\Delta_1=0$ and $\chi_3\neq0$,
$\Delta_3\neq0$ (consequently $\langle\mathbf S_i\cdot\mathbf
S_{i+1}\rangle=0$, $\langle\mathbf S_i\cdot\mathbf
S_{i+2}\rangle<0$).

{\subsection{ Competing orders and the phase diagram}}

It is known that the AFM Heisenberg model on triangular lattice
with $J_1>0, J_3=K=0$ has a $120^\circ$ ordered ground state (with
wave vector $(\frac{1}{3}, \frac{1}{3}, 0)$). When $J_1,J_3>0$,
the classical ground state is still ordered, but with an
incommensurate wave vector. The $K$ term gives rise to spin
nematic order through the identity\cite{AFQ,Nematic}
\begin{equation}  \label{nematic}
(\mathbf S_i\cdot\mathbf S_j)^2= Q_i^{\alpha\beta}
Q_j^{\alpha\beta},
\end{equation}
where $Q^{\alpha\beta}={1\over2}(S^\alpha S^\beta+S^\beta
S^\alpha)$ is the spin quadrupole tensor. To incorporate these
possibilities in our theory, we introduce additional decouplings
in our mean field decomposition.

To reduce the number of trial parameters in our calculation we
assume that the long-ranged magnetic order, if exist, is always
$120^\circ$ ordered in $H_{MF}$. To introduce the AFM order, we
divide the triangular lattice into three sublattices $u$,$v$ and
$w$ as shown in Fig. \ref{fig:pairing}(b). We assume without loss
of generality that the direction of long-ranged magnetic order $
\langle\mathbf S_{a}\rangle$ (here $a\in\{u,v,w\}$) is pointing
along the new basis axis $\tilde {\mathbf x}_{a}$ of the
$a$-sublattice (see Fig. \ref{fig:pairing}(b), i.e. ${\mathbf
M}_a=\langle \mathbf S_a\rangle=\langle \tilde
S^x_a\rangle\tilde{\mathbf x}_a=M\tilde {\mathbf x}_a$ in the new
reference frame and becomes an effective ferromagnetic order. The
operators $C$(in the old frame) and $\tilde C$(in the new frame)
obey the relations $C_u=\tilde C_u$, $ C_v=e^{-iS_z\theta}\tilde
C_v$ and $C_w= e^{iS_z\theta}\tilde C_w$, where $\theta=2\pi/3$.
Then Eq. (\ref{SS}) becomes $\mathbf S_{ai}\cdot\mathbf S_{bj}=
-:(\tilde\chi_{ai,bj}^\dagger\tilde\chi_{ai,bj}+\tilde
\Delta_{ai,bj}^\dagger\tilde\Delta_{ai,bj}):$, where $(i,j)$ and
$(a,b)$ are the site and sublattice indices respectively and
\begin{eqnarray}\label{chi'}
\tilde\chi_{ai,bj}&=&e^{-i\theta}\tilde c_{1ai}^\dagger \tilde
c_{1bj}+\tilde c_{0ai}^\dagger\tilde c_{0bj}+e^{i\theta} \tilde
c_{-1ai}^\dagger\tilde  c_{-1bj},  \nonumber\\
\tilde\Delta_{ai,bj}&=& e^{-i\theta}\tilde c_{-1ai}\tilde
c_{1bj}-\tilde c_{0ai}\tilde c_{0bj}+e^{i\theta}\tilde
c_{1ai}\tilde c_{-1bj},
\end{eqnarray}
for $(a,b)\in\{(u,v),(v,w),(w,u)\}$. Including the mean field
decoupling $\mathbf{S}_i\cdot\mathbf{S}_j\sim\langle\mathbf{S}_i
\rangle\cdot\mathbf{S}_j+\mathbf{S}_i\cdot\langle\mathbf{S}_j\rangle-
\langle\mathbf{S}_i\rangle\cdot\langle\mathbf{S}_j\rangle$ where
$\langle{\mathbf{S}}\rangle=M\tilde{\mathbf x}$, we obtain
\begin{eqnarray} \label{decouple}
 \mathbf S_{ai}\cdot\mathbf S_{bj}&\sim&-[(\chi\tilde\chi^\dagger_{ai,bj}+\Delta
 \tilde\Delta^\dagger_{ai,bj}-M\cos\theta\tilde S^x_{bj})+h.c.]\nonumber\\
 &&+\chi^2+\Delta^2-M^2\cos\theta,
\end{eqnarray}
where $\chi=\langle\tilde\chi_{ai,bj}\rangle$ and
$\Delta=\langle\tilde\Delta_{ai,bj}\rangle$.

We next consider the $K$ term. First we observe that the $K$ term
can be decoupled as $K(\mathbf S_i\cdot\mathbf S_j)^2\sim
K'\mathbf S_i\cdot\mathbf S_j$, where $K'=K\langle\mathbf
S_i\cdot\mathbf S_j\rangle$, and $\mathbf S_i\cdot\mathbf S_j$ can
be further decoupled as in (\ref{decouple}). This decoupling
renormalizes $J_1$. On the other hand, the $K$ term may give rise
to nematic order according to equation\ (\ref{nematic}) and a
corresponding mean field decoupling can be introduced in our
calculation with
\begin{eqnarray}\label{QQ}
(\mathbf S_i\cdot\mathbf S_j)^2\sim \langle Q_i^{\alpha\beta}
\rangle Q_j^{\alpha\beta}+Q_i^{\alpha\beta}\langle
Q_j^{\alpha\beta}\rangle-\langle Q_i^{\alpha\beta} \rangle \langle
Q_j^{\alpha\beta}\rangle.\\ \nonumber
\end{eqnarray}

\begin{figure}[hpbt]
\includegraphics[width=8.2cm]{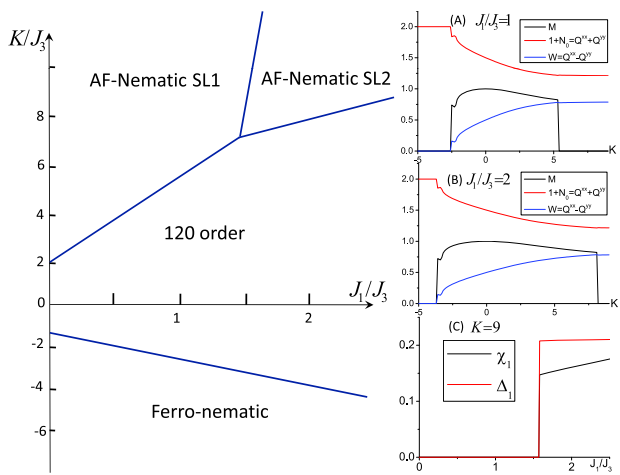}
\caption{(Color online) Phase diagram based on the mean field
theory. All the phase transitions are first order. The two spin
liquid phases are coexisting with $120^\circ$ anti-ferro nematic
order. Insets (A), (B) and (C) show the parameters change along
the lines of $J_1/J_3=1,2$ and $K=9$ respectively. }
\label{fig:phasediag}
\end{figure}

We shall assume that $\langle\tilde Q_i^{\alpha\beta}\rangle$ is
diagonalized in the new frame (hence the trial wave function has a
$120^\circ$ nematic order). So we have $\langle Q_i^{\alpha\beta}
\rangle Q_j^{\alpha\beta}\rightarrow\cos^2\theta\langle \tilde
Q_i^{xx} \rangle\tilde Q_j^{xx}+\sin^2\theta\langle \tilde
Q_i^{xx}\rangle \tilde Q_j^{yy}+\sin^2\theta\langle \tilde
Q_i^{yy}\rangle \tilde Q_j^{xx}+\cos^2\theta\langle\tilde Q_i^{yy}
\rangle\tilde Q_j^{yy}+\langle\tilde Q_i^{zz} \rangle\tilde
Q_j^{zz}$. It is easy to show that $\tilde Q^{xx}=\tilde
S_x^2={1\over2}[1+\tilde c_0^\dagger\tilde c_0+(\tilde c_1^\dagger
\tilde c_{-1}+ \tilde c_{-1}^\dagger \tilde c_1)]$, $\tilde
Q^{yy}=\tilde S_y^2={1\over2}[1+\tilde c_0^\dagger\tilde
c_0-(\tilde c_1^\dagger \tilde c_{-1}+\tilde c_{-1}^\dagger \tilde
c_1)]$ and $\tilde Q^{zz}=\tilde S_z^2=1-\tilde c_0^\dagger\tilde
c_0$. Putting together, we obtain our mean field decoupling
\begin{eqnarray}\label{K}
&K(\mathbf S_i\cdot\mathbf S_j)^2  \sim K'\mathbf S_i\cdot\mathbf
S_j+2K[({3\over2}N_0-{1\over2}) \tilde c_{0i}^\dagger\tilde
c_{0i}& \nonumber \\&\ \ \ \ \ \ \ \ \ \ \ \ \ -{1\over4}W(\tilde
c_{1i}^\dagger \tilde c_{-1i}+\tilde c_{-1i}^\dagger \tilde
c_{1i}) +{3\over4}-{1\over4} N_0],
\end{eqnarray}
where $N_0=1-\langle\tilde Q^{zz}\rangle=\langle\tilde
c_{0i}^\dagger \tilde c_{0i}\rangle$ and $W=\langle\tilde
Q^{xx}-\tilde Q^{yy}\rangle=\langle \tilde c_{1i}^\dagger \tilde
c_{-1i}+\tilde c_{-1i}^\dagger \tilde c_{1i}\rangle$ are two mean
field parameters representing nematic order. Notice that $N_0>1/3$
implies easy $\tilde x\tilde y$-plane anisotropy and nonzero $W$
indicates anisotropy of the quadrupole in $\tilde x\tilde
y$-plane. The total mean field Hamiltonian is thus
\begin{widetext}
\begin{eqnarray}\label{H_m}
H_{MF}&=&\sum_{\langle i,j\rangle}(-J_1+
K')\left[(\chi_1\tilde\chi^\dagger_{ij}+\Delta_1\tilde
\Delta^\dagger_{ij}-M\cos\theta\tilde
S^x_{i})+h.c.\right]-J_3\sum_{[i,j]} \left[(\chi_3\tilde
\chi^\dagger_{ij}
+\Delta_3\tilde\Delta^\dagger_{ij}-M\cos\theta\tilde S^x_{i})+h.c.\right] \nonumber\\
& & +KZ\sum_i\left[({3\over2}N_0-{1\over2})\tilde
c_{0i}^\dagger\tilde c_{0i}-{1\over4}W(\tilde c_{1i}^\dagger
\tilde c_{-1i}+\tilde c_{-1i}^\dagger \tilde
c_{1i})\right]+\lambda\sum_i(\tilde c_{1i}^\dagger \tilde
c_{1i}+\tilde c_{0i}^\dagger \tilde c_{0i}+\tilde
c_{-1i}^\dagger\tilde c_{-1i}).
\end{eqnarray}
\end{widetext}
The mean field Hamiltonian can be diagonalized straightforwardly
and the self-consistent equations for the mean field parameters
are similar to Eq.(\ref{MFeqs}) except the presence of three more
order parameters $M=\langle\tilde S^x\rangle, N_0=\langle \tilde
c_{0i}^\dagger\tilde c_{0i}\rangle$ and $W=\langle \tilde
c_{1i}^\dagger\tilde c_{-1i} +\tilde c_{-1i}^\dagger\tilde
c_{1i}\rangle$. We find more than one solutions to above
equations, and the one with lowest energy is chosen to be the
ground state. The phase diagram(Fig. \ref{fig:phasediag}) is
constructed by finding the mean-field ground states with different
parameters $K/J_3$ and $J_1/J_3$.

Transitions between different phases are found to be all first
order. The phase in the bottom of the phase diagram with negative
$K$ is an easy-plane ferro-nematic phase. In this phase, $N_0=1$
and $\chi_{1,3}=\Delta_{1,3}=M=W=0$. All spin correlations
$\langle\mathbf S_i\cdot\mathbf S_j\rangle=0$ vanish and
$\langle(\mathbf S_i\cdot\mathbf S_{i+1})^2\rangle=2$. The ground
state is a direct product state $\prod_i|\psi_i\rangle$, with
$S^{z}_i|\psi_i\rangle=0$. When $|K|$ becomes smaller, it goes
into the $120^\circ$ ordered phase, where $M\sim1$, $W\sim0.5$,
$\Delta_{1,3}=\chi_{1,3}=0$ and $\langle\mathbf S_i\cdot\mathbf
S_j\rangle\sim-0.5$ for NN and NNNN. When $K$ increases further,
there appear two nematic phases. These are anisotropic spin
phases with fermionic excitation similar to the two classes of
spin liquids we found when $K=0$ except that the spectrum is split
into three separate branches with two branches gapped. The new
feature here is that $W\neq0$, meaning that a $120^\circ$ AF
nematic order is built in (Here $120^\circ$ AF nematic order means
that the in-plane easy-axis of $Q^{\alpha\beta}$ form $120^\circ$
angle between any two neighboring sites), and the fermionic spinon
spectrum is modified. The ground state is no longer a spin
singlet.

It should be noted that our mean field ansatz (\ref{H_m}) is not
able to include several plausible states, like the magnetic
ordering at angles $\neq120^\circ$, or the $90^\circ$ AFN phase
proposed in\cite{AFQ}.
 Therefore, our mean field phase diagram should be
considered as suggestive only. A more accurate phase diagram can
be obtained only when the above plausible states are taken into
account and the energies are calculated more accurately from,
e.g., the Gutzwiller projected ground state wave function.
Nevertheless, our calculation shows the possible existence of spin
liquid states for spin systems with $S>1/2$. The spin liquid state
can be stabilized (at small $K$, without AF nematic ordering) by
the ring exchange interactions\cite{ring} which are not included
in our present study.

\section{Discussion and conclusion}
Before concluding this paper, we compare our theory with the
experiment on the recently discovered magnetic insulator $\mathrm
{NiGa_2S_4}$ \cite {science, PRBexp, PRBrapid,J1J3,AFQ,FQ,Qphase,
J1J3K, Vorex, SU(3)}. In this compound, $S=1$ $\mathrm{Ni}^{2+}$
ions form a layered triangular lattice with antiferromagnetic
(AFM) interaction. The system was found to be in a spin disordered
state at temperatures down to $0.35$K despite a Weiss temperature
$\theta_W\sim -80$K.
The $T^2$ temperature dependence of specific heat at low
temperature (below $10$K) indicates that spin excitation is
gapless while magnetic susceptibility approaches a constant below
$10$K\cite{science}. Several possible ground state have been
proposed and studied\cite{science, AFQ, FQ, Vorex, JPSJ}.
Here we propose that the $f$-wave spin liquid state we obtained is
a plausible ground state. In this case, nodal points appear in the
spectrum of low lying spin excitations, resulting in $T^2$
temperature dependence of specific heat which is consistent with
the experiment\cite{science}. The $f$-wave state also predicts
linear temperature dependence of spin susceptibility at low
temperature. However it should be noted that the existing sample
$\mathrm{NiGa_2S_4}$ is strongly disordered and is presumably in a
spin-glass state at low temperature. A clean sample is desired for
better characterization of the material.

Summarizing, in this paper we have generalized the fermionic
representation of $S=1/2$ spins to spins with arbitrary magnitude.
A mean field theory is developed for a $S=1$ spin model where
several spin liquid solutions are obtained. We have also obtained
a AF-nematic state with fermionic spinon excitations. Our approach
opens the possibility of constructing new classes of spin states
for systems with spin magnitude $S>1/2$.

We thank Prof. P. A. Lee and Prof. N. Nagaosa for valuable
suggestions and discussions. We also thank Mr. C. Chan for helpful
discussions. ZXL and TKN are supported by RGC grant of HKSAR.
YZ is supported by the Fundamental Research Funds for the Central Universities
in China.



\begin{thebibliography}{99}
\bibitem{footnote} We have also investigated $p$-wave states and found
that they break the rotation symmetry and are energetically higher
than the $p_x+ip_y$-state.
\end{thebibliography}
\end{document}